\documentclass[12pt]{iopart}
\usepackage{citesort,epsfig}
\advance\textwidth 3cm
\advance\textheight -1cm
\advance\topmargin 1cm
\begin{document}

\title{Specific protein-protein binding in
many-component mixtures of proteins}
\author{Richard P. Sear
}
\address{Department of Physics, University of Surrey,
Guildford, Surrey GU2 7XH, United Kingdom 
{\tt r.sear@surrey.ac.uk}}

\begin{abstract}
Proteins must bind to specific other proteins {\em in vivo}
in order to function. The proteins must bind only
to one or a few other proteins of the of order a thousand
proteins typically present {\em in vivo}.
Using a simple model of a protein,
specific binding in many component mixtures is studied. It
is found to be a demanding function in the sense that it demands
that the binding sites of the proteins be encoded by long sequences
of elements, and
the requirement for specific binding then strongly constrains these
sequences.
This is quantified by the capability of proteins of a given size
(sequence length), which is the maximum number of specific-binding
interactions possible
in a mixture. This calculation of the maximum number possible
is in the same spirit as the work
of Shannon and others on the maximum rate of
communication through noisy channels.
\end{abstract}
\maketitle

\section{Introduction}

Proteins must interact to function, and they must interact with very
high specificity in the highly complex mixtures that lie inside
cells. For example, the 3 proteins Cdk2, cyclin and PCNA
bind together in the nucleus to form a complex, which then goes on to interact
with high specificity 
with a few other proteins \cite{koundrioukoff00}.
(PCNA=Proliferating Cell Nuclear Antigen and
Cdk2=Cyclin dependent kinase; they are both involved in cell division but
also in other processes.)
Cdk2 must bind strongly to cyclin and to PCNA, i.e., the complex
must have a large equilibrium constant or equivalently a small dissociation
constant. But Cdk2 must {\em not}
bind to the of order one thousand other proteins
present i.e., the equilibrium constants with all these other proteins
must be small. This highly specific and strong binding in a mixture
of thousands of different molecules is a demanding requirement
and here we will try to quantify how demanding it is for a very
simple model of a protein. To do so, we will rely on the fact
that specific binding in a mixture is analogous to communication
through a noisy channel, which has been extensively studied,
in particular by Shannon \cite{shannon48,shannon} and the many
who have built on his work.
In communication through
a noisy channel reliable communication requires that all possible
messages be sufficiently distinct from each other even after
distortion by the noise. In specific
binding in a mixture, a protein binding site or patch on the
surface of a protein must be sufficiently distinct from all other
patches to allow one other patch to bind to it with an equilibrium
constant much larger than that of any other interaction of the patch.
The theory developed by Shannon has been enormously useful in
understanding communication systems and building efficient ones.
We hope that applying the theory to
protein-protein interactions in cells
will be similarly useful.

This analogy to communication
has been used for the similar problem of binding
between a protein and a specific DNA sequence, in particular by
Schneider \cite{schneider94,schneider86,schneider01,schneider91}.
See also \cite{gerland02}
for other related work. Similar approaches have also been used
extensively, see the review \cite{perelson97}, in studies of our
immune system, see \cite{thecell} for an introduction
to our immune system.
Studies of the immune system look at a large set
of immune-system proteins recognising a single foreign protein.
By contrast,
our situation, a model of the cytoplasm of a prokaryote
or a compartment of a eukaryote, involves not a single
protein binding to another but binding
between a very large number of pairs of proteins.
However, there are close analogies between the specific
binding required inside all cells and that required of immune-system
proteins, in particular in both cases proteins must avoid
binding to the wrong protein \cite{perelson79,perelson97,percus93}.
Inside cells, this would be all
proteins bar a protein's partner, whereas in the case of the immune
system, its proteins must not bind to other human proteins.

Our example of Cdk2 needing to bind to PCNA and to cyclin,
is the rule rather than the exception in eukaryote
cells \cite{thecell}. We merely chose a specific
example to make the discussion more concrete.
The cells rely totally on a complex network or
web of many specific interactions.
This large set of interactions is often called the interactome,
by analogy to the use of the word genome to denote the set of genes
of an organism. In the study of noisy channels, a fundamental
quantity of interest is the channel capacity
\cite{shannon48,shannon,mackay}:
the maximum number of bits of information that can be sent per second.
Shannon derived his famous channel-capacity theorem
\cite{shannon48,shannon}
which allows this channel capacity to be calculated. Here we
will calculate a related quantity for proteins,
which we will call the capability.
It is the maximum number of
specific interactions,
the maximum size of the interactome.

Protein interactions need to be understood to understand how cells
work, but also the specificity of interactions is crucial to drug
development. For example, Colas {\em et al.} \cite{colas96} used
a combinatorial-library approach to obtain peptide aptamers
that bind specifically to Cdk2.
Peptide aptamers are proteins that consist of a basic globular protein
to form a `scaffold' plus a region where the amino acids are allowed
to vary.
A library of these is generated and
the peptide aptamers whose variable regions bind most strongly
to the selected protein, here Cdk2, are selected.
To function effectively, the potential drug, here the peptide aptamer,
must interact in a highly specific way. If the requirement is to
interfere with Cdk2, then the drug should bind to Cdk2 but not to
the other thousands of proteins present. It is not sufficient that
it bind strongly to Cdk2 in a dilute solution {\em in vitro}.

In the next section we will specify our mixture that is a model
for the cytoplasm. It is an $N$ component mixture with highly
specific interactions between the components.
Section \ref{secmodel}
defines the model of a protein. It is a simple lattice
model in which each model protein has 6 binding sites or surface
patches that mediate the specific interactions.
Each patch is specified by a sequence of elements, and the interactions of
a pair of patches is determined by their sequences. The model is not
new, it was introduced in \cite{sear03b}, but there only
solubility, not specific binding was considered.
The capability, the maximum number of interactions of a given
high specificity, is
calculated in section \ref{seccap}, as a function of the length of these
sequences. The next section, section \ref{secres}, contains
some illustrative results.
The final section is a conclusion. An appendix contains a discussion
of Shannon's channel-capacity theorem and a brief analysis along
the lines of Schneider \cite{schneider94,schneider86}.

\section{Model of the cytoplasm of a cell}
\label{secmix}

Our picture of the mixture of proteins inside a cell is of
a large number of proteins, interacting via interactions of high
specificity \cite{thecell}. Our model of this is a mixture with $N$
distinct
patches on proteins engaging in $N/2$
pairwise interactions of high specificity.
These are not necessarily due to $N$ proteins.
$N$ is an even number.

In practice, in a cell the pairwise binding
interactions vary widely in strength, and some protein binding
sites, which we model by patches, bind to more
than one other protein.
Our model protein can be used to deal with
this variation but for simplicity we take all $N/2$ interactions to
be equivalent. Also, as the number of proteins increases
individual proteins become more dilute, this will tend to require
larger equilibrium constants for binding. We neglect this effect here.

So, in a mixture with $N$ binding patches,
each odd numbered patch $i$
is required to bind with the $(i+1)$th patch, with a large
equilibrium constant $K_b$. Now, the $i$th  ($i$ odd) and $(i+1)$th proteins
must bind only to each other, it must be specific, and
so in addition to binding to each other we require that
the equilibrium constant for binding of patch $i$
to any patch other than patch $i+1$ be less than $K_s\ll K_b$.
Similarly,
the equilibrium constant between patch $i+1$ and
any protein other than protein $i$ must also be less than $K_s$.
Thus if we denote the binding constant between proteins
$i$ and $j$ by $K_{ij}$ we require
\begin{eqnarray}
K_{ij}&>& K_b ~~~  |i-j|=1~~~~~\min\{i,j\}~~\mbox{odd}\nonumber\\
K_{ij}&<& K_s ~~~ \mbox{otherwise}
\label{func}
\end{eqnarray}

\begin{figure}[h]
\caption{
\lineskip 2pt
\lineskiplimit 2pt
Schematic representation of a model protein,
with the 3 visible patches represented by `barcodes': a sequence
of stripes, light for hydrophilic and dark for hydrophobic.
The model shown has $n_E=4$ elements of which 2 are hydrophobic (0)
and 2 are hydrophilic (1) in each case.
For example, the `barcode' of the front patch is $0101$.
\label{cube}
}
\vspace*{0.3in}
\begin{center}
\epsfig{file=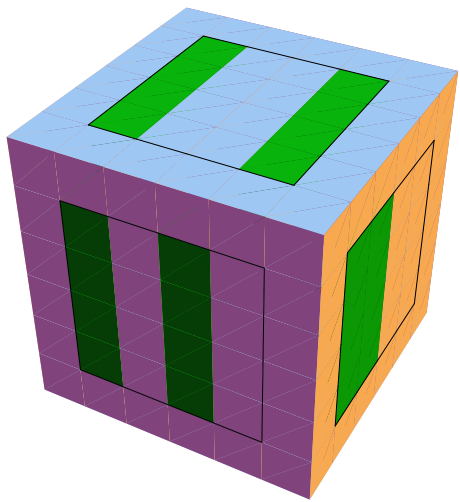,width=2.5in}
\end{center}
\end{figure}

\section{Model protein}
\label{secmodel}

Our model protein is as first defined in \cite{sear03b},
although there only solubility, not binding was considered, and
because of this
only patches with rather short sequences were considered.
The model protein is chosen to be as simple and as generic as possible,
while having interactions which are
mediated by surface patches whose interactions
are a function of the values of a sequence or string of elements.
Figure \ref{cube} is a schematic of the model. Note that
although the elements determine the interactions in our model and the
residues determine the interactions in a real protein, there is
no one-to-one relationship between one of our elements and an amino acid.

The model protein is a cube, with each of its 6 faces
having a single patch, for details see \cite{sear03b}.
The lattice
is cubic and each protein occupies 8 lattice sites arranged
2 by 2 by 2.
We make the model 2 sites across to reduce the
range of the attraction, which is 1 site, to half the diameter
of the hard core: see \cite{sear02} where an earlier
version of this model was defined and its behaviour compared
to that of real proteins.
The model proteins can rotate, and so
have 24 distinct orientations. Each of the 6 faces of the cube
has a patch.
If the faces of 2 proteins
are in contact
there is an energy of interaction between the 2 touching patches of the
2 proteins. By in contact we mean that
the faces must
overlap completely otherwise the energy of interaction is taken to be zero.
Also, the model is such that the energy of interaction between two
touching patches is a constant which does not change when the two
proteins are rotated about the axis joining their centres.

How a patch interacts
is specified by a
sequence or string of $n_E$ elements.
If a element has a value of 1 then
the element is said to be hydrophilic or polar, whereas if it has a value of 0
then it is hydrophobic.
A schematic of our model protein is shown in figure \ref{cube},
where the sequence is represented by a `barcode' with light
stripes indicating 1s and dark stripes indicating 0s.
The interaction energy, $u_{ij}$, of a pair
of touching patches, $i$ and $j$, is then given by
\begin{equation}
u_{ij}=-\epsilon\sum_{\alpha=1}^{n_E}\left(b^{(i)}_{\alpha}-1\right)
\left(b^{(j)}_{1+n_E-\alpha}-1\right),
\end{equation}
where $b^{(i)}_{\alpha}$ is element number $\alpha$ of patch $i$.
$\epsilon$ is the interaction energy of 2 elements that are hydrophobic.
We use energy units such that the thermal energy $k_BT=1$.

To calculate the interaction the string of elements
of one of the patches is reversed and then the energy is just
$\epsilon$ times the
sum of the number of pairs of corresponding elements where both
elements are 0, are hydrophobic. 
The only interaction is between 2 hydrophobic
elements; there is no hydrophobic-hydrophilic or
hydrophilic-hydrophilic interaction.
The reason one of the strings is reversed is that if this is not
done then the interaction between like patches, $j=i$, is just
$\epsilon$ times the number of 0s in the string. Then all patches
would stick to themselves.
Reversing
the strings removes this problem in a simple way.
Of course, the interactions form a symmetric square matrix,
$u_{ij}=u_{ji}$.
Thus, the binding site of
a protein is specified by giving values to the string
of $n_E$ elements, and so there are $2^{n_E}$ possible different patches.

The binding constant $K_{ij}$ of a patch is given by
\cite{sear03b}
\begin{equation}
K_{ij}=\left(1/2\right)^{\delta_{ij}}
\exp\left(u_{ij}\right)/6,
\label{kdef}
\end{equation}
where the first factor is a symmetry factor which
halves the equilibrium constant if $i=j$.
The factor of $1/6$
is a normalisation factor of $1/24$, from the 24 distinct
orientations, times the
4 possible rotations about the axis joining the centres of the proteins,
all of which allow binding.

So, our interaction is mediated by patches, of which we have
$N$ patches interacting in pairs. We neglect the other patches
on the surface of the proteins,
simply assuming that they are highly hydrophilic,
having few hydrophobic elements, and so can be neglected.
We do not need to specify the number of proteins $N_{PR}$, although
of course as each protein has 6 patches, we must have that
$N_{PR}\ge N/6$.

Although we are focusing on proteins, a mixture of particles that
satisfies equation (\ref{func}) will self-assemble in a controlled way,
the particles which have patch 1 will only stick to those that
have patch 2, those with patch 3 stick to those with patch 4 and so on.
We can even consider macroscopic objects with
interaction energies determined by sequences made of
large hydrophobic and hydrophilic stripes, e.g., a cube
just as in figure \ref{cube} and perhaps a few mm or cm across.
If these objects in water are agitated sufficiently strongly
to break the weak bonds, those with $K<K_s$, but not the stronger
bonds, those with $K>K_b$, then the cubes should self assemble.
For systems roughly equivalent to our model with small $n_E$,
this has been done in a series
of beautiful experiments by Whitesides and coworkers
\cite{whitesides02,bowden97}.

\section{Calculation of the capability}
\label{seccap}

As each patch must recognise a unique
other patch, each patch must have a unique sequence of
hydrophobic and hydrophilic elements.
As with $n_E$ elements there are $2^{n_E}$ different sequences
then immediately we know that $2^{n_E}\ge N$, or
$n_E\ge\log_2(N)$. This is an obvious lower bound on the number
of elements required for proteins to bind specifically to 1 other
protein from $N$.
It does not take into account our requirement,
equation \eref{func}. If we use all $2^{n_E}$ sequences then many
sequences in the mixture will differ from other sequences
present by the value of only 1 element.
Yet we require that a patch on a protein
bind very strongly to one other patch and very weakly to all
the others. For example, we require that patch 1 bind to patch
2 strongly but patch 3 very weakly or not at all.
Thus,
we require that $K_{12}> K_b$ and $K_{13}<K_s$. But if say
proteins 2 and 3 differ in only one element
then $K_{12}$ and $K_{13}$
can differ by a factor of at most $\exp(\epsilon)$. So, then
$K_{12}/K_{13}\le\exp(\epsilon)$, which is inconsistent with
our requirements for $K_{12}$ and $K_{13}$ unless $\epsilon$ is large.
In other words as the ratio $K_b/K_s$ is large,
in order satisfy our criterion
for functionality, equation \eref{func}, we cannot have
2 sequences present in which only 1 element is different. This
requirement to `space out' the patches
dramatically reduces the number of proteins we can simultaneously
use in a mixture.
We will need considerably more than $\log_2(N)$ elements
per protein patch to achieve
a functional mixture.

In fact this problem of a patch picking out one other patch
from a total of $N$ patches is analogous to problems that arise in
the study of communication through noisy channels.
There the receiver of a message must pick out which of the possible
messages they have received.
The study of communication channels
gave birth to what is now
often called information theory \cite{mackay}.
In the 1940s
Shannon wanted to know how many messages could be sent through
a given noisy channel during some time interval, with each message
being received correctly \cite{shannon48,shannon}.
Reference \cite{shannon48} is reprinted in \cite{shannon}.
The theory is statistical in nature, in the sense that it does not
consider a specific message but considers typical messages.
Our theory too is statistical, we will not consider a specific
mixture.

Now, we require that a protein bind to another with a large
equilibrium constant $K_b$. This will require that the
protein have a minimum of $B$ hydrophobic elements, where
\begin{equation}
B=\mbox{int}\left(\ln(6K_b)/\epsilon\right)+1,
\label{bdef}
\end{equation}
from equation \eref{kdef}. The function $\mbox{int}(x)$ yields the largest
integer less than $x$. It is required as the number of elements $B$ must
of course be an integer.
As we will see, the more hydrophobic
elements a patch has the more other patches it sticks to,
i.e., the more other possible patches have $K_{ij}>K_s$, and
so the number of hydrophobic elements should be kept to a minimum.
We demonstrate this at the end of this section.
As it cannot be less than $B$ while still binding strongly to one
other patch, the optimum number of hydrophobic elements is $B$.
From now on, we only consider patches with $B$ hydrophobic elements.
With this restriction the number of possible patches, $N_{POS}$, is
\begin{equation}
N_{POS}=
\begin{array}{c}
\mbox{Number of}\\
\mbox{proteins with $B$}\\
\mbox{hydrophobic elements}
\end{array}
=\frac{n_E!}{(n_E-B)!B!}.
\label{number}
\end{equation}
Actually, this neglects the
fact that some of these sequences are not actually
possible as they bind to themselves with $K_{ii}>K_s$, and
this makes equation \eref{number} a slight overestimate.
The mean number of interactions of a patch with itself is close to
$B^2/(2n_E)$. In the next section we will be considering patches where
$B^2/(2n_E)$ is fewer than the
number of hydrophobic interactions that make an interaction too sticky.
So we will use equation \eref{number} and
neglect the fact that a small
fraction of patches stick to themselves.


But any one of the possible proteins will stick to many
other proteins.
We use the term
`stick to' to mean bind with an equilibrium constant $K_{ij}>K_s$.
We need to calculate the number of proteins $j$
for which $K_{ij}>K_s$.
We denote this number by $N_S$.
If $S$ is the minimum number of elements for
which $K_{ij} > K_s$, then $S$ is given by
\begin{equation}
S=\mbox{int}\left(\ln(6K_s)/\epsilon\right)+1.
\label{sdef}
\end{equation}
Now, if $N_s$ is the number of patches with $B$ hydrophobic elements, $s$
of which interact with a fixed set of $B$ elements on another patch, then
the number of patches that stick
to a given patch, $N_S$, is
\begin{equation}
N_S=\sum_{s=S}^BN_s,
\label{equiv}
\end{equation}
because any patch with $s\ge S$ hydrophobic interactions contributes
to the number of patches that are too sticky.
Calculating $N_s$ is a simple
exercise in combinatorics.
Consider a patch with $B$ elements.
The number of ways, $N_s$, that $s$ elements can be situated such that
they interact with a fixed pattern of $B$ elements out of $n_E$ is
\begin{equation}
N_s=\frac{B!}{(B-s)!s!}\frac{(n_E-B)!}{(n_E-B-(B-s))!(B-s)!}.
\label{nstick}
\end{equation}

Having found expressions for the total number of patches, $N_{POS}$,
and the number each one sticks to, $N_S$, we proceed to see
how many specific interactions are possible. The first thing
to note is that once a patch is specified, so is its partner.
All $B$ of the partner's hydrophobic elements must be at the $B$ positions 
that interact with the hydrophobic elements of the first protein and so
there is no freedom in choosing its sequence of elements.
Thus, as there is a unique partner for each patch, we have
$N_{POS}/2$ binding pairs of patches. Actually
a very small fraction of patches bind to themselves, but we
neglect this.

So, we have $N_{POS}/2$ pairs. Every time we add one of these pairs
to the mixture, we have to discard approximately $2N_S$ of these pairs.
This figure of $2N_S$ is obtained by making two assumptions.
The first is that
each of the 2 patches of the pair eliminates $N_S$ pairs by
sticking to one of the partners --- we neglect sticking to more than
one patch of a binding pair. The second is that there is no overlap between the
sets of approximately $N_S$ pairs that each of the 2 patches of the pair
added sticks to. If every time we add a pair we have $2N_S$ fewer
pairs then clearly we run out of pairs after we have
added $(N_{POS}/2)/(2N_S)$ pairs. We name
the maximum number of patches
that can bind together in pairs in a mixture,
the capability and denote it by $C$. For our model it is
\begin{equation}
C(n_E,\epsilon)=\frac{N_{POS}}{2N_S}.
\label{capability}
\end{equation}
The maximum number of high specificity bonds is of course half the
capability $C$. In information theory, the fundamental quantity
is the capacity $C_{Sh}$. In units of bits per bit sent, it
is related to our capability by
$C_{Sh}=\log_2(C)/n_E$ \cite{shannon48,shannon,mackay}.
In communication the analogue of our capability $C$ increases exponentially
so it is convenient to use not the capability itself but its logarithm
the capacity. As we will see our $C$ does not increase exponentially
with $n_E$ and so it is more convenient to work with it directly.
In an appendix we discuss the similarities and differences to
communication.

To derive equation \eref{capability} we assumed that
$n$ patches stick to $nN_S$ others, neglecting any overlap
between the $N_S$ patches that stick to one patch and the $N_S$
patches that stick to another.
A simplifying
assumption that means that equation \eref{capability} will tend to
underestimate the true capability.


For their pioneering study of the immune system,
Perelson and Oster \cite{perelson79}
introduced the idea of `shape space' of proteins where if a protein
from the immune system is near in this space to a foreign protein,
it will bind to, recognise, the foreign protein.
Our model proteins too have a corresponding space and the capability
we have calculated
essentially corresponds to the number of proteins
required to fill this space. By fill we mean have so many
proteins that any extra one of the $N_{POS}$ patches with $B$
hydrophobic elements that is added is likely to
stick to one of the patches in the mixture.
For an immune system to function the space should be overfilled
to avoid any foreign protein escaping detection
\cite{perelson79,perelson97,percus93},
it is an advantage for more than one immune-system protein to bind
to a foreign protein,
whereas for specific
binding the space must be underfilled.

When the
first term in equation \eref{equiv} for $N_S$ is the dominant one,
we can approximate $N_S$ as
\begin{eqnarray}
N_S&\simeq&
\frac{B!}{(B-S)!S!}\frac{(n_E-B)!}{(n_E-B-(B-S))!(B-S)!}\nonumber\\
&\simeq& \frac{B!n_E^{B-S}}{S!(B-S)!^2},
\label{nsapp}
\end{eqnarray}
where in the second line we simplified the expression assuming
$n_E-B$ is not too small.
Making similar approximations for $N_{POS}$ we get
$N_{POS}\simeq n_E^{B}/B!$ and hence for the capability $C$,
\begin{eqnarray}
C&\simeq&\frac{1}{2}\frac{n_E^{B}}{B!}
\frac{S!(B-S)!^2}{B!n_E^{B-S}}=
\frac{(B-S)!^2S!}{2B!^2}n_E^S.
\label{capp}
\end{eqnarray}
Also, note that from this approximate expression
it is clear that to maximise the capability the number of hydrophobic
elements, $B$ in equation (\ref{capp}) should be as small as possible.
In equation (\ref{capp}), the $B$ dependence is
$C\sim 1/[B(B-1)\ldots(B-S+1)]$ and this is a monotonically decreasing
function of increasing $B$.
Finally, as promised $C$ does
not increase exponentially with the number of elements $n_E$.
We should note that Shannon's channel-capacity theorem
\cite{shannon48,shannon} implies
that there potentially exists a model in which $C$ does increase
exponentially. So real proteins may have a $C$ that increases
exponentially with the number of amino acids in a binding patch,
however in our model the increase is sub-exponential.

\begin{figure}
\caption{
\lineskip 2pt
\lineskiplimit 2pt
Plots of the capability $C$ as a function of the number of elements $n_E$.
For both curves $\epsilon=1$ and the maximum permitted
stickiness is $K_s=100$ ($S=7$). For the solid curve the binding
strength is $K_b=10^6$ ($B=16$) and for the dashed curve
$K_b=10^8$ ($B=21$).
The dotted curves just below the solid and dashed curves are
the results of the approximation
equation \eref{capp} for $C$.
\label{nprot}
}
\vspace*{0.3in}
\begin{center}
\epsfig{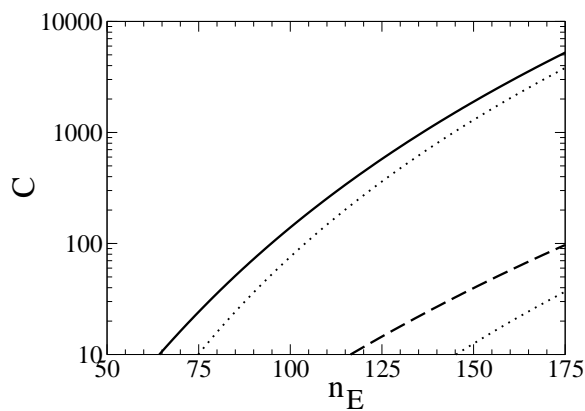}
\end{center}
\end{figure}

\section{Results for the capability}
\label{secres}

The binding strength of protein interactions are often given
in terms of the dissociation constant,
typically with Molar (M) units \cite{thecell}. 1 Molar is
$6\times 10^{23}$ molecules per litre, which is $0.6$ molecules/nm$^3$.
The dissociation constant, $K_d$, for $A$ and $B$ binding
together is, by definition,
$K_d=c_Ac_B/c_{AB}$, where $c_A$, $c_B$ and $c_{AB}$ are the
Molar concentrations of free (unbound) $A$, free $B$ and bound
$AB$ dimers, respectively \cite{thecell}. It is therefore just one
over the equilibrium constant for binding.
Dissociation constants of protein interactions vary over orders
of magnitude. They can easily be of order $\mu$M and
in some cases values as low as $10^{-16}$M have been measured.
For example, some {\em E.~coli} can produce a protein called E9 (it
is one of a family of related proteins)
which is
highly toxic, it is an antibiotic. E9 is neutralised by another
protein, Im9, which binds to it to form a heterodimer
\cite{wallis95,wallis98,kuhlmann00}. At low salt {\em in vitro}
the dissociation constant for the binding of Im9 to E9 is
close to $10^{-16}$M. 
In $250$mM sodium chloride solution the dissociation constant is
a mere  $10^{-14}$M: the binding has a significant
electrostatic component. This is a remarkably strong and specific
binding, it must be specific as it must avoid binding {\em in vivo}
to proteins other than E9.

E9 and Im9 are a somewhat extreme example, to get a feel for
more typical values, consider
a bacterium such as {\em E.~coli}. It has a cytoplasm
with a volume around 1$\mu$m$^3$. So, one protein
in this volume has a concentration of $10^{-9}$nm$^{-3}$
or about 1nM.
If there are only say 10
copies of $A$ and $B$ then for a significant fraction of these copies
to bind to each other, then a dissociation constant around
a $10$nM is required.
If there
are around 1000 copies of the proteins $A$ and $B$ in each
{\em E.~coli} cell then a dissociation constant of around 1$\mu$M
is enough to ensure roughly equal numbers of the free $A$ and $B$
and of $A$ and $B$ bound into dimers.

Our proteins are 2 lattice sites across and
our equilibrium constants $K_{ij}$ are in units of the volume
of a lattice site.
Proteins are a few nms across and so each lattice site corresponds to
a volume about
1 to 2nms across. Taking a lattice site to be 1nm across,
and noting that 1Molar is close to 1 per nm$^3$ we have that
$K_{ij}=1$ is approximately equivalent to an equilibrium constant of 1M$^{-1}$.
Thus to obtain a dissociation constant of 1$\mu$M we need a
$K_{ij}\approx10^6$, and so need to set $K_b=10^6$.

In figure \ref{nprot} we have plotted the capability $C$ for $K_b=10^6$
(solid curve) and $10^8$ (dashed curve). In both cases we
insisted that the equilibrium constant between patches not binding
to each other be less than $K_s=100$.
A $K_b=10^8$ is approximately equivalent
to a dissociation constant of 10nM.
We took the
energy of interaction between hydrophobic elements $\epsilon=1$.
Then for $K_b=10^6$, $B=16$ and,
for $K_b=10^8$, $B=21$, and with $K_s=100$,
$S=7$ elements is too sticky, $K_{ij}>100$. Also, the dotted curves
are the result of equation \eref{capp}.

\begin{figure}
\caption{
\lineskip 2pt
\lineskiplimit 2pt
Plots of the number of possible proteins to which a protein
binds, $N_S$, as a function of the number of elements $n_E$.
For both curves $\epsilon=1$ and the maximum permitted
stickiness is $K_s=100$ ($S=7$). For the solid curve the binding
strength is $K_B=10^6$ ($B=16$) and for the dashed curve
$K_B=10^8$ ($B=21$).
\label{npfan}
}
\vspace*{0.3in}
\begin{center}
\epsfig{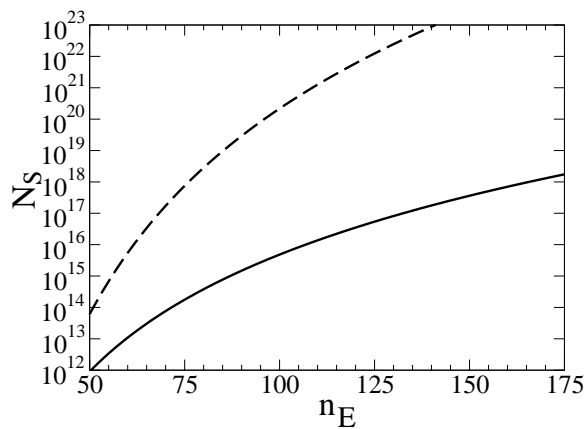}
\end{center}
\end{figure}

Of course the capability $C$ increases as $n_E$ increases but
for the stronger binding, even for $n_E=150$, $C=40$.
This is highly inefficient, the restriction
to be functional means that we need $150$ elements for a
mixture with 40 specific binding interactions,
whereas if we just needed 40 different patches
then $\log_240=5.3$ elements would have been enough.
However, each patch of a protein is required to bind to 1 other
patch with a very large equilibrium constant, and with
equilibrium constants at least 6 orders of magnitude smaller
to any of the other 39 patches. This requirement
is a very stringent one.

The inefficiency can be clearly seen in figure \ref{npfan} where
we have plotted the number of possible patches to which a patch
sticks. We see that if we insist on specific binding of strength
$K_b=10^8$ (the dashed curve)
then we require $B=21$ hydrophobic elements for bonding of this strength
but then when $n_E=150$, a patch with 21 hydrophobic elements sticks
to approximately $2.8\times 10^{23}$ other possible proteins.
The equilibrium constant between a single patch and of order $10^{23}$
others is greater than $K_s=100$. Thus to make a mixture that is
functional according to our criteria, for every specific-binding
patch we add to
the mixture the number of possible patches that we can no longer
add to the mixture increases by $10^{23}$.

The same thing must apply to real proteins {\em in vivo}, although
proteins are much more complex than our, deliberately very simple,
model and so it is not possible to quantify $N_S$
for real proteins. However, we can return
to our example of E9 and Im9
\cite{wallis95,wallis98,kuhlmann00}.
A number of single-residue mutants have been made for
approximately 30 of the residues
of Im9 suspected of contributing to the E9-Im9 binding. The changes
in the dissociation constant in each case were measured, and they
cover a large range \cite{wallis98} but many of them resulted in
only modest increases in the dissociation constant. Thus, many
sequences with these mutations are not viable as part 
of the amino-acid sequences of proteins. We cannot have,
for example, an enzyme that
has a part of its amino-acid sequence close to that of Im9
as the enzyme will bind
to E9. Many more mutants, mutants with 2, 3, etc. mutations,
will also bind strongly to E9 and in general this will be inappropriate.
So functional proteins cannot include these amino-acid sequences,
thus reducing the number of amino-acid sequences which are functional
in the mixture of proteins in the cytoplasm of {\em E.~coli}
when E9 is present.


\section{Conclusion}

The mixture of proteins inside cells is not simply an inert mixture
like, for example, the complex mixture of hydrocarbons in crude oil.
It has evolved
to be functional: the proteins do things. In particular, many
proteins must bind strongly to a specific patch or patches on one or
more other proteins. Here, we studied a very simple model protein,
introduced in \cite{sear03b}.
We borrowed ideas from the theory of communication, and calculated
the maximum number of specific binding interactions
that can function together in a mixture. We called it the
capability $C$.
It
is roughly the number of possible different patches, $N_{POS}$,
over the number of patches that stick to a given patch and
so cannot be in the same mixture, $N_S$, $C=N_{POS}/2N_S$,
The capability $C$ of the constituent proteins limits
the size of the interactome. The
interactome is a network or web of many specific interactions, and it
can have at most $C/2$ links.
A network with many links requires that the
proteins of which it is composed must have large patches, patches with
a large number of elements.

Schneider has applied information theory to analysing the DNA
sequences of sequences of base pairs to which proteins bind
with high specificity \cite{schneider94,schneider86}. We briefly
apply an analogous analysis to our model in the appendix.
Work on the size of the regions on foreign proteins
that are recognised by the immune-system proteins is very
similar in approach to that used here \cite{perelson97}.
For example, Percus {\em et al.}'s \cite{percus93}
study of how demanding is the requirement
that our immune-system proteins differentiate between foreign and
human proteins, is analogous to our study of how demanding
is the requirement that a protein bind to only one other
patch in a mixture. However, we do use a concrete although
highly simplified, model of a protein unlike work on the immune
system which has used rather abstract models of recognition.
Indeed, it might be useful to use our concrete model of a protein
in future studies of the immune system.


The more specific the binding, the larger $K_b/K_s$ at fixed $K_s$,
the more elements $n_E$ we need to achieve a given capability $C$,
see figure \ref{nprot}.
Now,
although there is no one-to-one correspondence between our
hydrophobic/hydrophilic elements and the residues of a real protein,
we expect that increasing $n_E$
corresponds in real proteins to increasing the size
of the protein.
But ideally, proteins should be as small as possible, they will then
consume less material (amino acids), diffuse faster, and take
up less volume.
For these reasons we might want $n_E$ to be as small
as possible and so there is a trade off between the size and
complexity needed for specific binding and the problems of being large.
This trade off involves not only the properties of an individual
protein but also the global properties of the mixture, e.g.,
the number of components.
Eukaryote
cells are compartmentalised \cite{thecell} and
each compartment will have only a subset of the total number of proteins.
A smaller number of proteins with higher concentrations
in a compartment allows strong specific binding at smaller values of $n_E$.

Very recent work on the specificity of protein-protein binding has been
done by
Havranek and Harbury \cite{havranek03}, and by Shifman and Mayo
\cite{shifman03}. They looked at specificity but in both cases
they used a combination of a much more detailed model
and experiments.
In particular, Havranek and Harbury considered 2 proteins
and constrained each protein to: i)
to dimerise with itself, ii) not to bind to the
other protein, iii) not to aggregate.
They found that proteins have to
be specifically engineered not to bind to other proteins and not
to aggregate,
as well as
to bind to their binding partner (themself in this case).
The detailed
nature of their model restricted them to a mixture of 2 proteins. It
also precluded the exhaustive enumeration of all possible
proteins and elimination of {\em all} of the proteins that bind
to proteins other than their partners, that we performed here.
Following the approach taken by Shannon \cite{shannon,shannon48} we also
did not consider a specific mixture, as Havranek and Harbury did,
but calculated a limit on what the mixture could do, here the limit
on the number of interaction of a given specificity.
However, theirs and our approaches are related. We both consider
how specificity is achieved and in doing so explicitly consider the binding
of a protein to the wrong other protein or site, that must be
prevented. This avoidance of an undesired state is referred
to as negative design
\cite{richardson02,wang02,doye03}.

Future work could also go in a number of other directions.
The concepts introduced
here of a capability $C$ for specific binding of a protein
could be applied to other models of proteins.
Also, the proteins in
a cell are the product of evolution of course, and it would
be interesting to study the constraints on evolution of proteins due
to the requirement of specific binding.
Very recent work \cite{fraser02,teichmann02,fraser03}
on large data sets of interactions, partial
interactomes,
has found a negative correlation between
the rate of evolution of the amino-acid sequence of a protein
and the number of specific interactions it participates in.
Although there is some debate over this finding
\cite{fraser03,jordan03,bloom03}.
This work relies on recent work characterising very large numbers
of interactions, particularly in yeast \cite{uetz00,ito00}.
Also see the references of
\cite{fraser02,teichmann02} for earlier work on individual proteins.
Within our model, if a protein patch is part of a mixture near
the capability of its proteins, then not only would many mutations
be deleterious because they weakened its binding to its partner,
many mutations would be deleterious because they caused the patch to
stick to other patches in the mixture. The mutation rate
of an individual protein will in general
depend on the sequences of elements of {\em all} the other patches,
not just that of its partner.



It is a pleasure to acknowledge that this work started
with inspiring discussions with D. Frenkel. I would also
like to acknowledge discussions with J. Cuesta, and
J. Doye and A. Louis for introducing me to negative design.
Finally, I would like to thank
an anonymous reviewer for drawing my attention to the connection
to the work of Whitesides.
This work was supported by the Wellcome Trust (069242).

\section*{Appendix}

In our notation, the channel-capacity theorem of Shannon states
that in the $n_E\to\infty$ limit, we can achieve a capability
$C$ such that $\log_2(C)/n_E\to O(1)$, with fixed $K_s$ and
the specificity ratio $K_b/K_s$ arbitrarily large. For our
model protein $C$ does not increase exponentially with $n_E$
and so in the $n_E\to\infty$ limit, $\log_2(C)/n_E\to 0$.
Thus the channel-capacity theorem
does not apply to our model of a protein.
Note that this does not rule out a different type of model
protein for which the channel capacity theorem holds, or that real proteins
might obey it.

Shannon's derivation of the channel-capacity theorem
(theorem 11 of \cite{shannon48,shannon})
breaks down when applied to our proteins
because he assumes, correctly
for communication, that the entropy per bit sent of the strings is some
constant which he calls $H$. Here the entropy per element $s$, is
\[
s=-\left(\frac{B}{n_E}\right)\ln\left(\frac{B}{n_E}\right)
-\left(\frac{n_E-B}{n_E}\right)\ln\left(\frac{n_E-B}{n_E}\right)
\]
which is the logarithm of equation (\ref{number}) over $n_E$.
At fixed $\epsilon$, $B$ is fixed and so as $n_E$ is increased,
$s$ decreases. In order to avoid the patches becoming very sticky
the number of hydrophobic elements must be minimised and this makes
$s$ decrease with increasing $n_E$ which in turn dramatically
reduces the capability.
This feature is present in our model but not
in communication systems. It is also not present in the more abstract
approach to applying information theory to high specificity binding
of Schneider \cite{schneider91,schneider94}.
It is possible that it is simply an artifact
of our simple model, but other related models do show a similar
behaviour \cite{unpub}. Another perspective on this is to note that
to make $s$ not a function of $n_E$, $B$ should not be a constant, it
should be $fn_E$, with $f<1$ a constant.
But then the average interaction energy between 2 patches
is $\epsilon f^2n_E/2$: which increases with $n_E$. In essence the
problem is that the larger the number of elements of the patch the larger
the interaction energy is. It is possible that a similar problem
also occurs in real proteins, where if a large patch on the surface
of a protein is required to achieve binding of high affinity it
is difficult to avoid this large patch interacting strongly with
many other proteins.
The analogue of the $\log(N_S)/n_E$ in communication,
called the equivocation, is also a constant, unlike  $\log(N_S)/n_E$.

Finally, it should be noted that $sn_E$ is simply the entropy
associated with the space of proteins that have the right number of
hydrophobic elements. This is the space consisting of $N_{POS}$ proteins.
It is not an information in the sense that the information of binding
sites on DNA evaluated by Schneider \cite{schneider94,schneider86} is an
information. He evaluates the information about a DNA sequence that
you obtain if you are told that the sequence is the binding site
for say the protein LacI \cite{schneider86}. This information,
call it $I_s$, is, by definition,
the difference between the entropy of the DNA
sequence ($2\times\mbox{number of base pairs}$ if the 4 base pairs
all occur with probability $0.25$) if you know nothing about it
and the entropy of sequences that LacI binds too. We can evaluate
a similar information, again call it $I_s$, for our model proteins.
Let us consider protein $i$, with $i$ odd so that it binds
to protein $i+1$.
$I_s$ is the information about the sequence of protein $i$ that
you obtain if you are told what the sequence is of protein $i+1$.
In section \ref{seccap} when we calculated the capability
$C$ we insisted on proteins having $B$ hydrophobic elements and then
once the sequence of protein $i+1$ is fixed, so is that of protein $i$
and so the entropy of sequence $i$ is then 0 once the sequence
of $i+1$ is known.  Then $I_s=sn_E$.
But $I_s$ need not always equal $sn_E$. In a mixture
of proteins with interactions coded by $n_E$ elements,
with $N$ below $C(n_E)$ more than one sequence $i$ will bind to $i+1$
and also satisfy the requirement not to stick to the other $N-1$ proteins.
The number of hydrophobic elements will not always be $B$.
Then the entropy of protein $i$ once the sequence of protein $i+1$
has been specified will be greater than 0 and $I_s$ will
not equal the entropy of the space of possible proteins.


\end{document}